\documentclass[a4paper,fleqn,usenatbib]{mnras}
\usepackage{soul}
\usepackage{newtxtext,newtxmath}
\usepackage[T1]{fontenc}
\usepackage{ae,aecompl}

\usepackage{deluxetable}

\usepackage{graphicx}	
\usepackage{amsmath}	
\usepackage{amssymb}	
\let\oldAA\AA
\renewcommand{\AA}{\text{\normalfont\oldAA}}

\usepackage{natbib}
\usepackage{ulem}
\usepackage{subfig}
\usepackage{float}
\usepackage[usenames,dvipsnames]{xcolor}
\usepackage{multirow}
\usepackage{footnote}
\usepackage[bottom]{footmisc}
\usepackage{ulem}
\usepackage{xltabular}
\usepackage{longtable}
\usepackage{lipsum} 
\usepackage{pdflscape}
\usepackage[export]{adjustbox}
\usepackage{ragged2e}

\title[UV, Optical and NIR Spectroscopy of SU Lyn]{UV spectroscopy confirms  SU Lyn to be a symbiotic star}

\author[Kumar et al.]{	
	Vipin Kumar$^{1,2}$\thanks{vipin@prl.res.in},
	Mudit K. Srivastava$^{1}$\thanks{mudit@prl.res.in},
	Dipankar P.K. Banerjee$^{1}$
	and Vishal Joshi$^{1}$	 
	\\
	$^{1}$Astronomy \& Astrophysics Division, Physical Research Laboratory, Ahmedabad 380009, India\\
	$^{2}$Indian Institute of Technology, Gandhinagar, 382335, India\\
}

\date{Accepted XXX. Received YYY; in original form ZZZ}

\pubyear{2020}

\begin{document}
	\label{firstpage}
	\pagerange{\pageref{firstpage}--\pageref{lastpage}}
	\maketitle

\begin{abstract}
SU Lyn, a star that ostensibly appears to be an unremarkable late M type giant, has recently been proposed to be a symbiotic star largely based on its hard X-ray properties. The star does not display, in low-resolution optical spectra,  the high excitation lines typically seen in the spectra of symbiotic stars. In the present work, UV, optical, and near-infrared observations are presented, aimed at exploring and strengthening the proposed symbiotic classification for this star. Our Far-UV 1300-1800$\AA$ spectrum of SU Lyn, obtained with the ASTROSAT mission's UVIT payload, shows emission lines of Si IV, C IV, OIII and N III in a spectrum typical of symbiotic stars. The UV spectrum robustly confirms SU Lyn's symbiotic nature.  The detection of high excitation lines in a high-resolution optical spectrum further consolidates its symbiotic nature. As is being recognized, the potential existence of other similar symbiotic systems could significantly impact the census of symbiotic stars in the galaxy.

\end{abstract}



 

\section{Introduction}
\label{sec-Intro}
Symbiotic stars are binary stellar systems consisting of a white dwarf (WD)/neutron star primary and a giant companion. These systems have traditionally been identified by the presence of strong emission lines of high excitation species in their optical spectra (e.g. \cite{Kenyon1986}). In addition, they often show two rather "exotic" lines at 6825$\AA$ and 7082$\AA$ which have been identified due to Raman scattering of the OVI 1032$\AA$, 1038$\AA$ resonance lines by neutral hydrogen \citep{Schmid1989}. The presence of the aforementioned emission lines has thus always been the conventional route for identifying such systems. Breaking this mould, SU Lyn - an infrared-bright (K{$_ {s}$} = 1.618) but otherwise unremarkable M type star, has recently been proposed to be a symbiotic star despite showing no obvious emission lines in its low-resolution optical spectra. SU Lyn was discovered in hard X-rays in the Swift/BAT hard X-ray all-sky survey by \cite{Mukai2016}.  These observations, coupled with a NuSTAR spectrum obtained by \cite{Oliveira2018},  showed that the X-ray emission could be modelled to arise from hot ionized plasma at $T$  $\sim$ 2x10$^{8}$K. The high temperature suggested an origin in a boundary layer between an accretion disk (AD) and a hot component. In short, the discovery of the hard x-ray emission indicated that SU Lyn had a hot component that was accreting matter. Based on the UV to X-ray flux ratio, the hot component in SU Lyn was inferred to be a WD rather than a neutron star. The distance to the system was estimated to be 640 $\pm$ 100 pc and the spectral type of the secondary was determined as M5.8III \citep{Mukai2016}. Complementing these observations, high-resolution optical observations revealed hitherto undetected emission lines in the visible spectra. The lines were weak, hence eluded detection in low-resolution spectra. Thus, the X-ray and the high-resolution optical spectrum taken together, implied that SU Lyn had the necessary properties of a symbiotic star. 
\par 
It is recognized (\cite{Luna2013, Mukai2016} and references therein) that symbiotics, based on ultraviolet(UV)-X-ray photometric/spectral observations, can be classified into (1.) shell-burning systems - powered by the processes of nuclear fusion as well as accretion and (2.) non-burning systems - powered by only the accretion process. The weakness of the visible emission lines and the rapid variability in the UV \citep{Mukai2016, Oliveira2018} suggested SU Lyn as a non-burning system, powered purely by accretion process. The lack of shell burning leads SU Lyn to have very weak symbiotic signatures in the optical and the observed UV luminosity predominantly originating from the accretion disc.
\par 
Given the lack of signatures in the optical and predominance of UV radiation in non-burning systems, it is expected that UV spectroscopy could be a powerful tool to probe the dynamics of such systems. The UV/optical spectra of well-established symbiotic stars show a complex blend of lines, e.g. forbidden, permitted and intercombination lines excited through a variety of atomic processes \citep{Meier1994}. These lines are suggested to originate from different regions of the symbiotic system. The lines of high ionization potential (IP) species, e.g. [Ne V], [Fe VI] etc. clearly demonstrate the existence of low-density regions of thin plasma heated and ionized from the UV radiation of the hot components. These lines presumably originate from the winds of WD and AD. The lines of lower IPs (e.g. N II, O I, O II, O III etc.) are most likely formed in the upper atmosphere of the red giant, or in the extended surrounding nebula (see \cite{Kenyon1984,Mikolajewska1989, Nussbaumer1989b} and references therein).
\par 
The UV emission lines (e.g [C IV] 1907,1909$\AA$, [N III] 1749$\AA$, [N IV] 1485$\AA$, [O IV] 1401$\AA$, Si III 1883,1892$\AA$) were found to be extremely helpful for emission-line diagnostics. Di-electric recombination lines from C II, C III, C IV, N III, N IV, N V, O III, O IV, Si III, Si IV etc. were found useful to estimate the electron temperature and densities of the nebula \citep{Hayes1986, Nussbaumer1987}. The UV spectra were also useful to determine the individual element abundances, their locations from the hot source and the temperature of the hot source \citep{Nussbaumer1989b}. However, the scenario is complex and detailed models (e.g. \cite{Vogel1994}), which depend on many input parameters, are needed to reproduce the strengths of the emission lines as well as the underlying continuum.

\par 
In the present work, optical, NIR and UV spectroscopy of SU Lyn are presented. The nature of the UV spectrum robustly entrenches SU Lyn as a symbiotic system. The high-resolution (R$\sim$30,000) optical spectrum detects several emission lines typical of symbiotic stars which are, however, not seen in the low-resolution (R$\sim$500-2000) data. A 0.8-2.4$\mu$m NIR spectrum at R $\sim$ 1000 of SU Lyn is also presented here in which we detect no emission lines, but interestingly HeI 1.083$\mu$m is seen in absorption. NIR spectra of two symbiotic stars are shown for comparison, to indicate the emission lines that could potentially be present.

\section{Observations and Data Analysis}
\label{sec-Obs}
\subsection{UV observations}
\label{subsec-UVObs}

The UV observations were done with the Ultra-Violet Imaging Telescope (UVIT) on the ASTROSAT satellite which consists of two telescopes of 38 cm diameter, one optimized for the Far UV (FUV, 1300-1800 $\AA$) and the other for the Near-UV (NUV, 2000-3000 $\AA$). The details can be found in \cite{Subramaniam2016,Tandon2017,Tandon2017b,Tandon2020}. The UV data presented here were obtained against proposals IDs AO4-026, and AO5-144 (observation IDs $A04\_026T01\_9000001712$ and $A05\_144T02\_9000003176$) wherein photometry was proposed to cover both the FUV and NUV filter bands. However, for certain technical reasons,  observations were executed only in the NUV bands in the NUVB15, NUVB4 and Silica filters (see table-note of Table~\ref{table-obs}). The slit-less FUV grism spectra were obtained in AO5 proposal cycle. The science ready level-2 data was provided by UVIT payload operation centre (POC) team through the ASTROSAT data archive. The reconstructed images and spectra are corrected for distortion, flat-field illumination, spacecraft drift and other effects as discussed in \cite{Tandon2017b, Tandon2020}. Aperture photometry, reduction of the spectra, wavelength and flux calibrations were done following prescribed procedures given in \cite{Tandon2017, Tandon2020}. Given the common pointing of ASTROSAT instruments, SU Lyn was simultaneously also observed in three of its X-ray instruments namely, LAXPC, SXT and CZTI. However, these data will be investigated separately.

\begin{table}
	\centering
	\caption{Log of Observations}
	\begin{tabular}{llll}
		
		\hline
		\hline		
		Instrument $\&$ Mode  &   Date (UT) &	Wavelength  & Integration\\
		(Facility) &             &  range (\AA) & Time (sec) \\
		\hline
		{\bf UV}               &             &               &     \\
		UVIT-Photometry$^a$             &  2017-11-22.26 & 2000-3000 & 1385,509,  \\
		(Astrosat)       &                 &               &    1853$^b$    \\		
		UVIT-Spectroscopy$^c$        &  2019-09-21.66  & 1300-1800 & 9542  \\
		(Astrosat)       &            &               &            \\
		\hline
		{\bf Optical}          &            &               &      \\
		HESP-Spectroscopy &  2018-01-09.81 & 3750-10500& 5400  \\
		R$\sim$30000, (HCT)$^d$         &                  &           &\\
		MFOSC-P Spectroscopy      &  2019-12-13.77$^e$  & 4500-8500 & Various   \\
		R$\sim$500-2000, (Mt. Abu)        &           &          &\\		
		\hline
		{\bf Near-Infrared}        &              &           &       \\
		NICS$^f$-Spectroscopy      &  2017-12-25.78 & 8400-24000& 380   \\
		R$\sim$1000, (Mt. Abu)  &&&\\	
		\hline				
		\hline
		
	\end{tabular}
	\label{table-obs}
	\begin{list}{}{}
		\item (a): The NUV telescope has 4 narrow band filters viz.  NUVB15($\lambda_{centre}$ 2196\AA, width 270\AA), NUVB13 (2447\AA, 280\AA), NUVB4(2632\AA, 275\AA), NUVN2(2792\AA, 90\AA) and two broad band Silica (2418\AA, 785\AA) filters. (b): In the NUVB15, NUVB4 and Silica filters respectively (c) : UVIT is  equipped with low resolution grisms (400 line pairs/mm) which provide a low resolution spectrum in slitless mode viz. $\sim$17$\AA$ in second order for the FUV (which was used here).  (d)  : Hanle echelle spectrograph of the Himalaya Chandra  Telescope  (e) : Several spectra were collected between 2019-2020. The one listed in this Table is used in this work. (f) : Near-IR Camera Spectrograph
	\end{list}
\end{table}

\begin{figure}
	\includegraphics[width=0.47\textwidth]{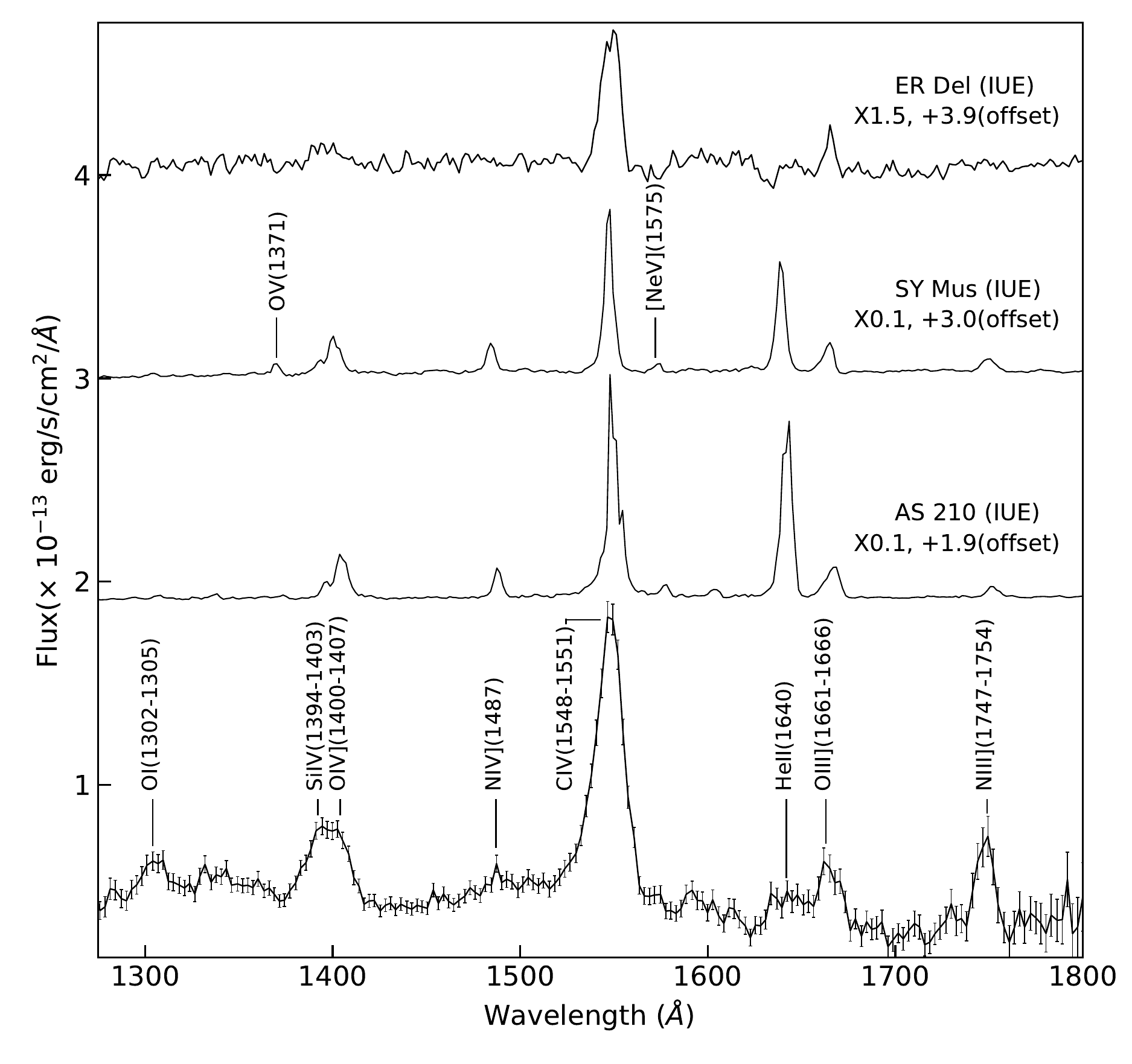}	
	\caption{The UVIT spectrum of SU Lyn with the lines identified. Archival IUE spectra of three symbiotics ER Del, SY Mus and AS 210, are also shown for comparison.}
	\label{fig-UVSpec}
\end{figure}

\subsection{ Optical and NIR (OIR) observations}
\label{subsec-OIRObs}

OIR spectroscopic observations were done with the PRL 1.2m Mt. Abu Telescope using the Near-Infrared Camera and Spectrometer (NICS) for the NIR 0.8-2.4 $\mu$m region and the Mt. Abu Faint Object Spectrograph and Camera (MFOSC-P) for the optical region \citep{Srivastava2018}. The details of the observations are presented in Table~\ref{table-obs}. Spectra have been reduced as per standard procedures described in several studies done earlier, e.g. \cite{Rajpurohit2020, Banerjee2018, Srivastava2016}. A high-resolution echelle spectrum covering 3750-10500$\AA$ in 61 orders at R$\sim$30000, was also obtained using the Hanle Echelle Spectrograph (HESP) on the 2m Himalayan Chandra Telescope (HCT). The pipeline reduced data was used for the analysis; further details of HESP can be found at https://www.iiap.res.in/hesp/.

\section{Results and Discussion}
\label{sec-Results}
\subsection{UV Spectroscopy and photometry}
\label{subsec-UVSpec}

The FUV spectrum of SU Lyn is shown in Fig~\ref{fig-UVSpec} with the lines marked. It shows the characteristic spectra of symbiotic stars, many of which are observed by the  International Ultraviolet Explorer (IUE) and reproduced for example in \cite{Michalitsianos1982} and in the catalogue of \cite{Meier1994}. For comparison, we also present in Fig~\ref{fig-UVSpec},  the spectra of three symbiotics viz. SY Mus, AS 210 and ER Del which are discussed shortly. To paraphrase \cite{Meier1994}, the UV spectrum of a symbiotic usually consists of a great range of excitation lines ranging from relatively low-temperature species (e.g., O I 1302-1305$\AA$, C II 1334, 1335$\AA$); strong resonance emission lines (e.g. N V 1238$\AA$, 1242$\AA$, C IV 1548, 1551$\AA$) and other high-ionization lines like He II 1640$\AA$. Diagnostic lines for studying the ambient nebular conditions is principally through the inter-combination lines (e.g. O IV$]$ 1397-1407$\AA$, N IV$]$ 1487$\AA$ etc.).


\begin{figure}
	\includegraphics[width=0.47\textwidth]{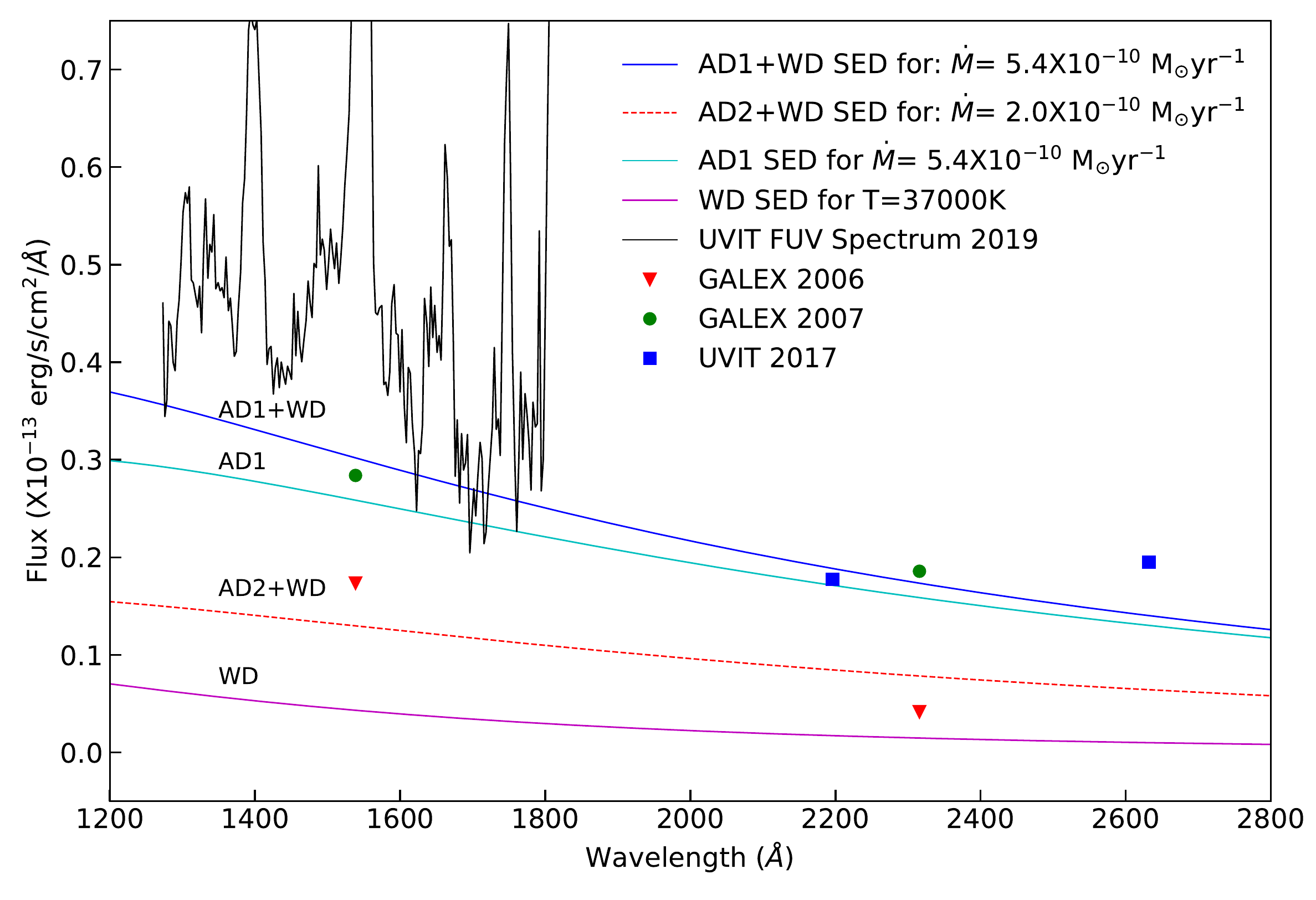}
	\caption{ Model fits to the UV spectrum and photometry using the combined spectral energy distribution (SED) of a WD of mass 0.85M$_{\odot}$ at 37,000K and an AD accreting mass at different rates as indicated in the inset. Most of the GALEX and UVIT data are reasonably reproduced using mass accretion rates of $\sim$ $10^{-10} M_{\odot}$ per year. The GALEX magnitudes in the NUV and FUV filters are 17.42$\pm$0.02 and 16.64$\pm$0.03 for Dec 2006 respectively; and 15.8$\pm$0.02 and 16.11$\pm$0.03 for Jan 2007 respectively. UVIT magnitudes/fluxes are given in the texts.}
	\label{fig-Model}
\end{figure}

\begin{figure}
	\includegraphics[width=0.47\textwidth]{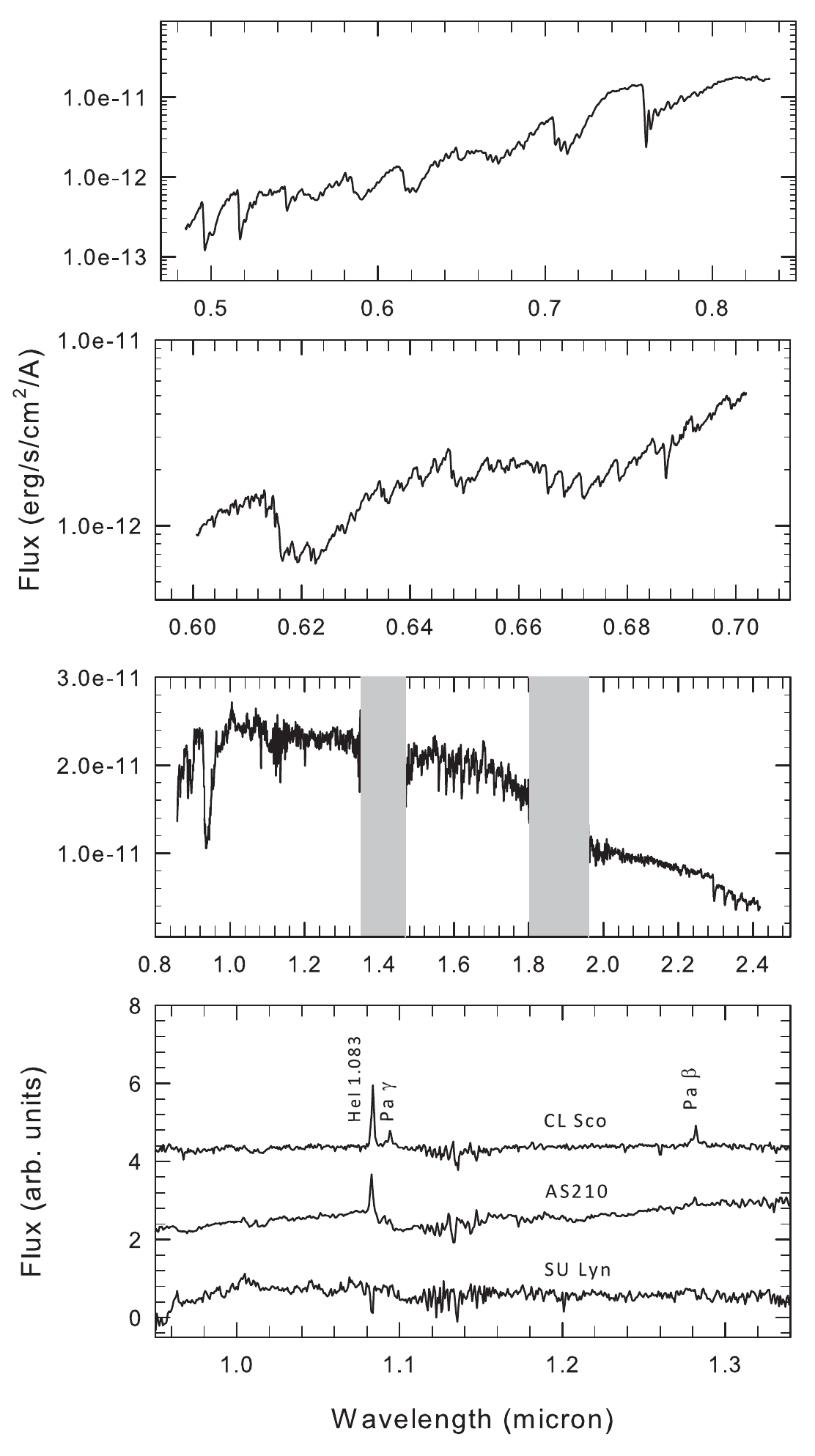}
	\caption{Top two panels show low-resolution optical spectra (at R$\sim$500 and 2000 respectively; epoch 2019 December 13) showing that emission lines are too faint to be seen at low resolution. The third panel shows the NIR spectrum, which is devoid of emission lines but HeI 1.083 $\mu$m is seen in absorption. The bottom panel presents an expanded view of the J band, which shows no emission lines in contrast to two other symbiotic stars CL Sco and AS210 whose spectra were obtained at the same resolution. The optical spectra have been flux-calibrated by matching their flux levels with the NIR spectrum at their common end regions (at $\sim$8400-8500$\AA$). The NIR spectrum has been anchored using the 2MASS $H$ band flux (H = 1.92 magnitudes). The spectra are not dereddened.}
	\label{fig-OptSpec}
\end{figure}


The lower spectral resolution of UVIT ($\sim$17$\AA$), as compared to other UV probes like IUE ($\sim$6$\AA$) or high-resolution GHRS/HST spectra, does not allow us to resolve components of some of the inter-combination lines to allow estimating density and temperature parameters (e.g. \cite{Keenan2002}), as discussed in Section~\ref{sec-Intro}. However, what we do see for certain is that He II 1640$\AA$ (the highest excitation line seen in our spectrum) is weak which likely implies that the temperature of the photo-ionizing source is not too high. In comparison, SY Mus is a shell-burning type of symbiotic system consisting of a hot $T=105000 K$ WD with a total luminosity $L=1600 L_\odot$. He II 1640$\AA$ line is among the brightest lines in its IUE spectrum, while emissions of [Ne V] 1575$\AA$ emission (IP=97.12 eV) and O V 1371$\AA$ (IP=77.41eV) has also been detected (\cite{Pereira1995} and reference therein). The optical and UV spectra of AS 210 are rich in emission features and displays high ionization lines, e.g. [Ne V] 3426$\AA$ \citep{Gutierrez-Moreno1996, Munari2002}. We wish to draw the reader’s attention towards the symbiotic system ER Del, which has been classified as a $\delta$-type system showing evidence of boundary layer X-ray emission \citep{Luna2013}. Its optical spectrum \citep{Munari2002} show only prominent H-$\alpha$ emission and IUE UV spectrum is devoid of high-ionization lines (even HeII 1640$\AA$ is very weak or absent). These features make ER Del a system closely resembling the characteristics of SU Lyn, a non-burning system.
\par 
The weakness of He II 1640$\AA$, in conjunction with the results from X-ray studies that strongly suggest that SU Lyn is a disk-dominated system, makes us propose that the central WD (which lacks shell-burning) is not the dominant source of the hard ionizing radiation in SU Lyn. Rather, photo-ionization likely occurs predominantly through the UV flux emitted by an AD which we propose is present (the presence of an AD is similarly proposed by \cite{Mukai2016} and \cite{Oliveira2018}). We thus explored whether the observed FUV continuum can be simulated with a simple model consisting of a WD and a steady-state, time-independent AD. The AD is approximated by a multi-colour black body with a radial temperature profile given by the standard disk model \citep{Pringle1981}, while the WD is considered to be a black body source. The results of the simulations are shown in Fig~\ref{fig-Model} against our UVIT spectrum and NUV photometry points. GALEX photometry points for 2006 and 2007 are also shown as given in \cite{Mukai2016}. All the photometry data and spectrum have been dereddened using the \cite{Cardelli1989} extinction law with $E(B-V) = 0.07$ as estimated by \cite{Mukai2016}.
\par 
Our UVIT photometry yielded magnitudes of 16.02$\pm$0.02 and 15.19$\pm$0.01 in the NUVB15 and NUVB4 filters with corresponding fluxes of $9.44\times10^{-15}$ and $1.28\times10^{-14}$ ergs/s/cm{$^ {-2}$}/$\AA$ respectively, before reddening correction. In the broadband NUV silica filter, we estimate a  magnitude of 16.25$\pm$0.01, but this observation may be affected by saturation effects as discussed in \cite{Srivastava2009}, and hence is not being used in our analysis. It may also be noted that significant fluctuation in the UV flux is seen between the GALEX and UVIT observations. SU Lyn is known to display considerable UV variability, both at sub-minute scale \citep{Mukai2016} and also on a longer time-scale. \cite{Oliveira2018} interpreted the UV variation between 2015 and 2016 to be a consequence of a large drop by almost 90$\%$ in the accretion rate. This aspect is discussed again in the next sub-section.
\par 
We attempted to reproduce the UV continuum through visual inspection of the simulated continuum. The mass of WD and the accretion rate were varied in the range of $0.7-1.0$ $M_\odot$  and $1.0-9.0 \times 10^{-10} M_\odot$ per year respectively, as per the estimates suggested by \cite{Mukai2016} and \cite{Oliveira2018}. The choice of these ranges then constrained the allowed temperature values of the WD. In order to reproduce appropriate UV continuum for SU Lyn UVIT spectrum and NUV photometry points, we tried WD temperature value in the range of $\sim30000-45000 K$. Through visual inspection, we found a suitable fit for the WD mass of $0.85 M_\odot$ and temperature of $37000K$. A wider range of temperatures may be allowed by the data and also be scientifically reasonable. However, our aim here is to provide a simple model till a higher-resolution UV spectrum is collected which we intend to obtain. As shown in Fig~\ref{fig-Model}, the continuum of the observed FUV spectrum is reasonably well reproduced using an accretion rate of $5.4\times10^{-10} M_{\odot}$ per year. The GALEX data of both the 2006 and 2007 epochs are fairly well fitted by using mass accretion rates of $5.4\times10^{-10} M_{\odot}$ per year and $2.0\times10^{-10} M_{\odot}$ per year respectively. The necessity of having to use different accretion rates, to fit the UV fluxes at different epochs, strongly suggests that it is the AD which predominantly powers the UV part of the spectrum and which is responsible for the variability seen in UV bands. It may also be pointed out that a very high flux ($\sim10^{-13}$ ergs/s/cm$^{2}$/$\AA$ was observed by Swift-UVOT in 2015 in the UVM2 band \citep{Mukai2016}. These data are not plotted in Fig~\ref{fig-Model} since the numerical values of the Swift fluxes are not given in \cite{Mukai2016}. However, our simulations indicate that such a level of the NUV continuum can be reproduced with a mass accretion rate of $\sim6.5\times10^{-9} M_{\odot}$ per year. This is not entirely inconsistent with the estimate of $1.6\times10^{-9} M_{\odot}$ per year, adopted by \cite{Mukai2016} to explain their observed UV luminosity with 1 M$_\odot$ WD.
\par 
In the simulations, the WD luminosity is estimated to be 0.16$L_{\odot}$  while the disk luminosity is significantly higher at 0.66$L_{\odot}$ with its flux peaking at $\sim1100\AA$. It thus suggests that the WD lacks the flux, in contrast to shell-burning symbiotics, to photo-ionize the wind of the red giant thereby giving rise to the characteristic intense emission line spectra of symbiotics. On the other hand, the more luminous AD could be the dominant photo-ionizing source. Further, the combined SED of the WD and the AD shows that the flux drops to insignificantly low levels by $\sim$ 220$\AA$ which corresponds to $\sim$ 56.4 eV. This could explain why no emission lines are observed with associated ionization potentials larger than 54 eV. However, we caution that our model is simplistic and we plan to obtain a higher resolution UV spectrum to enable more rigorous modelling (e.g. \cite{Nussbaumer1989b,Proga1996,Mikolajewska2006}) to account for the strengths of both the observed UV emission lines and the faint lines seen in the high-resolution optical spectra.
\par 
The absence of O V 1371$\AA$ line in the FUV spectrum also rules out the possibility of O VI being present. The O V 1371$\AA$ line is predominantly produced by dielectronic recombination from O VI \citep{Espey1995}. This is consistent with the absence of 6825$\AA$ and 7082$\AA$ features, as they are caused by the Raman scattering of O VI 1032, 1038$\AA$ resonance transitions. It is worth noting that it is mostly symbiotic stars of the burning type that are known to show these Raman scattered lines \citep{Munari2019}.

\subsection{Optical and Infrared spectroscopy}
\label{subsec-OIRSpec}

\begin{figure}
	\centering
	\includegraphics[width=0.47\textwidth]{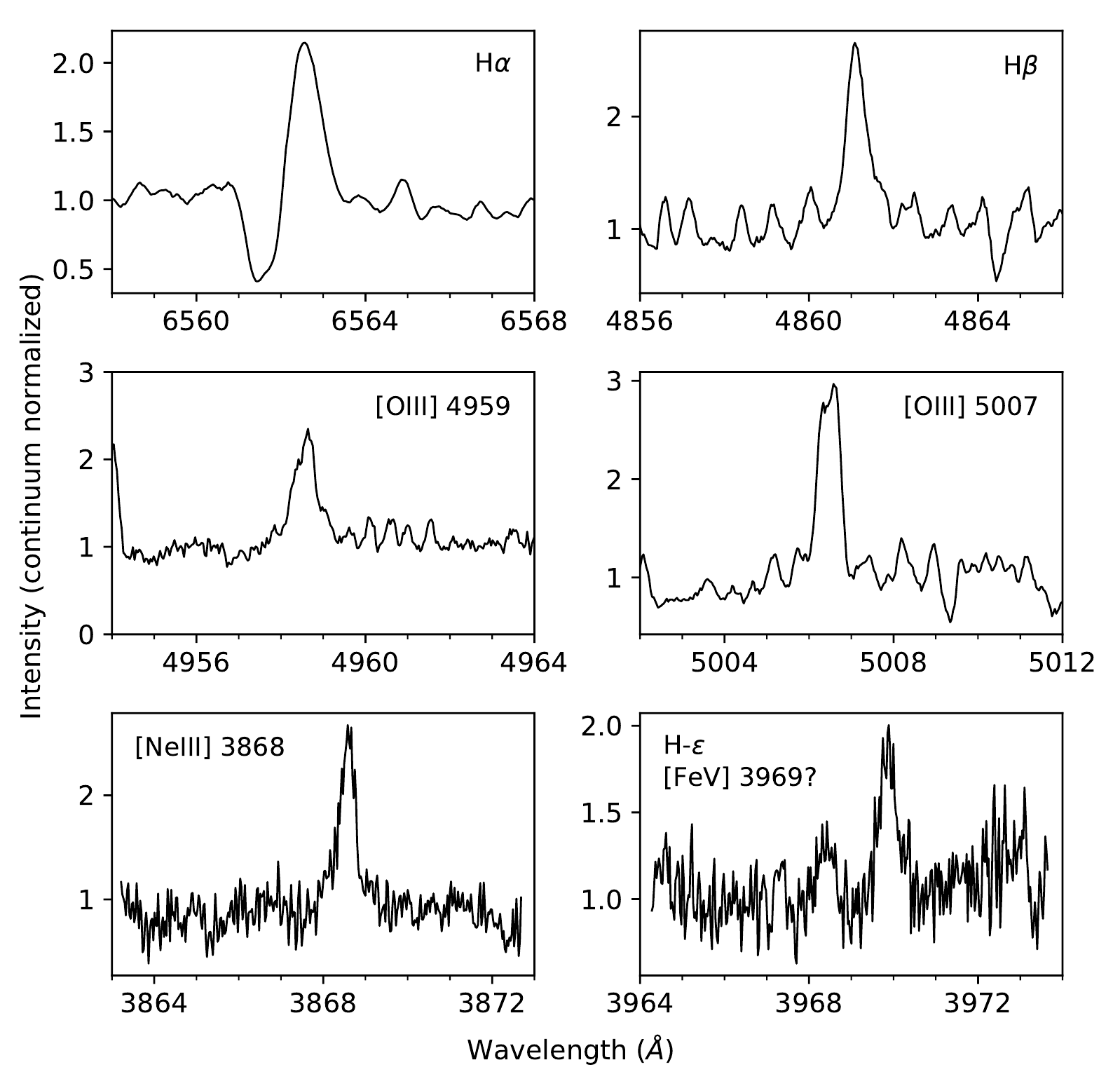}
	\caption{Emission lines seen in the high resolution spectrum of SU Lyn.}
	\label{fig-HespSpec}
\end{figure}

The top two panels of Fig~\ref{fig-OptSpec} show low-resolution optical spectra of SU Lyn wherein no emission lines are seen. However in the echelle spectra several emission lines are detected, some are shown in Fig~\ref{fig-HespSpec}, including Balmer H$\alpha$ to H$\epsilon$. For reasons that are unclear, He I lines are absent. We are also unsure whether He II 4686$\AA$ is present; there is a faint feature at that position. We also observe a feature at the position of [Fe V] 3969.4Å (IP = 54.8 eV), but we consider it more likely to be H-$\epsilon$ 3970$\AA$ because of a better wavelength match of the line centres. Thus, He II 1640$\AA$ (IP = 54.4 eV) represents the highest stages of ionization associated with the lines observed here. The higher critical densities  of $\sim$ $10^{9}- 10^{10} cm^{-3}$ for some of the intercombination lines (N III$]$ 1747,1754$\AA$; O III$]$ 1661,1666$\AA$) versus the relatively lower critical densities of $\sim$ $10^{5}- 10^{7} cm^{-3}$ for the forbidden lines (such as [OIII] 4959$\AA$, 5007$\AA$ and [Ne III] 3868$\AA$) in SU Lyn spectra, suggests (after \cite{Mikolajewska1989}) that the intercombination lines are formed closer to the hot component, while most of the forbidden lines occur in the extended nebula.
\par 
The H$\alpha$ line shows a  P Cygni profile which is relatively rare in symbiotics. \cite{Ivison1994} have presented multi-epoch high-resolution echelle spectroscopy of 35 symbiotics of which 27 have double-peaked H-$\alpha$ profiles. Only in TX CVn, a deep P Cygni profile has been observed. They propose that the predominance of reversed profiles with the blue component stronger, supports the idea that the majority of the structured  Balmer lines arise from self-absorption by neutral gas in the cool companion's wind. \cite{Munari2019} has shown six high-resolution H-$\alpha$ profiles of SU Lyn between October 2015 to December 2017 of which P Cygni profiles are seen on two epochs. Since P Cygni profiles arise in an accelerated region of the wind and also need large absorption column densities, we propose that there may be episodic phases of enhanced mass loss from the cool companion which manifest in P Cygni profiles. A consequence of the enhanced mass loss episodes would be corresponding increases in the accretion rate, causing enhancements in the UV luminosity.This speculative interpretation could explain the large UV variability seen in the star.
\par 
The NIR spectrum showed in Fig~\ref{fig-OptSpec} (third panel from top) is dominated by the red giant; the spectrum is typical of a late M type star. Both first and second overtone bands of CO are strongly prominent in the H and K bands, respectively. No evidence of dust is seen in our data, consistent with the lack of an IR excess in its SED plotted using  WISE, AKARI and IRAS data. No evidence of emission lines is seen, consistent with the findings of the low-resolution optical spectra. For comparison, the new, unpublished spectra of two symbiotics CL Sco and AS 210, obtained from Mt. Abu, are plotted (bottom panel) showing the lines that may be expected to be seen. A puzzling feature is that HeI 1.083$\mu$m is seen in absorption. The fact that HeII lines are seen in emission (1640$\AA$ and perhaps 4686$\AA$) whereas HeI is in absorption indicates a different site of origin for these lines. CL Sco has been classified as a low ionization symbiotic system \citep{Kenyon1984}. Though its optical spectrum in outburst \citep{Munari2002} shows bright emission line typical of a symbiotic system, the IUE UV spectrum showed a very weak He II 1640$\AA$, absence of N V 1239$\AA$, and a modest C IV emission. Moreover, the He II 4686$\AA$ was not present in the optical spectrum recorded close to the epoch of IUE observations \citep{Michalitsianos1982}.
\par 
To summarize, the UV spectrum presented here, complemented by optical and NIR spectra, provides direct and robust evidence that SU Lyn is a non-burning symbiotic star. The presence of emission lines in low-resolution optical spectra has been the traditional way to identify and discover symbiotics. But such a technique will fail in the case of SU Lyn-type of objects thereby underestimating the census of symbiotic systems in the galaxy. The number of known symbiotics stands at $\sim$400 at present but is rapidly expanding as a consequence of surveys of the Galactic Bulge and Plane (\cite{Munari2019} and references therein). It is estimated that the total Galactic population of symbiotic stars is approximately a factor of 1000 more than the known number \citep{Munari1992}. But this number may need to be substantially revised if SU Lyn type systems are included.
\section{acknowledgments}
We thank the reviewer for the useful suggestions. MKS expresses deep thanks to Shyam N. Tandon and Ranjeev Misra (IUCAA, India) for discussions related to this work. VK thanks PRL for his Ph.D. research fellowship. DPKB is supported by a CSIR Emeritus Scientist grant. We thank the staff of IAO, Hanle and CREST, Hosakote (both operated by the Indian Institute of Astrophysics, Bengaluru), that made these observations possible. This publication uses the data from the AstroSat mission of the Indian Space Research Organisation (ISRO), archived at the Indian Space Science Data Centre (ISSDC). UVIT was developed by a collaboration between IIA, IUCAA, TIFR, and several centres of ISRO from India, and CSA from Canada. \\
{\bf DATA AVAILABILITY}: The UVIT data will be available on the Astrosat archive at the end of the proprietary period. For the other data, the authors may be contacted.


\bibliography{BibList}{}
\bibliographystyle{mnras}

\end{document}